# Equation-Free Multiscale Computations in Social Networks: from Agent-based Modelling to Coarse-grained Stability and Bifurcation Analysis

**A.C. Tsoumanis[1], C. I. Siettos[2*], I.G. Kevrekidis[3] and G.V. Bafas[1]**

[1]School of Chemical Engineering, National Technical University of Athens,
GR 157 80, Athens, Greece
[2]School of Applied Mathematics and Physical Sciences, National Technical University of Athens,
GR 157 80, Athens, Greece
[3]Dept. of Chemical Engineering, Program in Computational and Applied Mathematics and
Dept. of Mathematics, Princeton University, Princeton, NJ 08544, USA

**Abstract**
We focus on the "trijunction" between multiscale computations, bifurcation theory and social networks. In particular we address how the "Equation-Free" approach, a recently developed computational framework, can be exploited to systematically extract coarse-grained, emergent dynamical information by bridging detailed, agent-based models of social interactions on networks, with macroscopic, systems-level, continuum numerical analysis tools. For our illustrations we use a simple dynamic agent-based model describing the propagation of information between individuals interacting under mimesis in a social network with private and public information. We describe the rules governing the evolution of the agents' emotional state dynamics and discover, through simulation, multiple stable stationary states as a function of the network topology. Using the "Equation-Free" approach we track the dependence of these stationary solutions on network parameters and quantify their stability in the form of coarse-grained bifurcation diagrams.

## 1. Introduction

Agent-based models constitute the state-of-the-art in contemporary computer simulation of many complex systems ranging from ecology [1-3] and epidemiology [4-6] to economics and financial markets [7-11], and from traffic and supply-chain networks [12-16] to biology [17,18] and physiology [19,20].

On the other hand, networks are identified as a key feature of the structure of many real-world systems and recent years' research on the subject is a part of a broader trend towards research on the dynamics of complex systems [21-27]. Within this context there has been an immense interest in studying social dynamics using detailed agent-based modelling on networks as they pertain to many complex systems such as epidemiological [28,29,5,30,31], economic and financial [32, 8, 33, 34, 9, 35-41], opinion formation and voter dynamics [42-48], culture and language dynamics [49-53], traffic and crowd flow design and management [54-57], diffusion of news and innovations [58-60]. The rapid technological and theoretical progress has dramatically enhanced our knowledge in modelling, simulating and analyzing such complex systems and we are getting better on it.

---
[*] Corresponding author

However due to the effectively complex, nonlinear, stochastic nature of such systems, their behavior at the macroscopic level–where usually the extensive analysis, design and optimization requirements arise- cannot be-most of the times-accurately modeled and analyzed in a straightforward manner. While one can try to write down a general master equation to describe the probabilistic time evolution of the macroscopic quantities, major problems arise in trying to find the closures required to bridge the gap between the micro-scale of the individual-based interactions on the network and the macro-scale at which the systems-level analysis and design are required [61, 46]. This lack of accurate, explicit macroscopic closures poses a major obstacle to the performance of important systems-level computational tasks, which rely on the availability of good low-order closed models in terms of a few macroscopic (coarse-grained) variables.

To date what is usually done with very detailed multi-agent-based descriptions of social networks is simple simulation: set up many initial conditions, for each initial condition create a large enough number of ensemble realizations, probably change some of the rules and then run the detailed dynamics for a long time to investigate how things like stock exchange indices or the spread rate of news change with parameters. However this approach is both time consuming and inappropriate for systems-level tasks such as stability, bifurcation analysis or control design and optimization since efficient well-established computational tools for such analysis and design tasks require the availability of coarse-grained models in the form of ordinary and/or (integro)-partial differential-algebraic (PDAEs) equations in a closed form.

In this work we show how the so-called "Equation-Free" approach for multiscale computations [62-68] can be exploited to systematically analyze the macroscopic behavior of agent-based simulations. For our illustrations we use a simple model describing the emotional state formation and propagation process under private and public information on a social network. Actually the model is a simple caricature of a society whose members are interested in some specific matter (e.g. a local election between two opponents, the trend to invest in a stock fund etc). We start with presenting how two important elements of the "Equation-Free' framework, namely the lifting and restricting operators, can be constructed to deal with certain features of the networks topology. We then construct the coarse-grained bifurcation diagrams with respect to network degree distribution and analyze the stability of the identified stationary solutions branches through coarse-grained eigen-computations.

The paper is organized as follows: in Section 2 we give a short description of the Equation-Free approach and the procedure for deriving the lifting and restricting operators. In Section 3 we describe the agent-based model as well as the algorithm we used to create random connected networks with degree sequences sampled from a specific probability distribution. More details are given in the appendix. In Section 4 we present and discuss the simulation results and finally we conclude in Section 5.

## 2. The Equation-Free Framework for Complex/ Multiscale Computations

The concept of the *coarse timestepper* is the key element in the "Equation-Free" framework [62-68], an efficient methodology for extracting systems-level information from microscopic (detailed, fine scale, atomistic, agent-based) simulations. For the approach to be applicable we must know –or have good reason to believe- that predictive coarse-grained models of the detailed process can be written in terms of (can "close" with) a reduced number of "coarse-grained variables" or "observables". Knowing such good variables through experience with the problem or through data-mining techniques [69, 70] is an important prerequisite. One might hypothesize, for example, that the average state (over several network realizations and over all nodes of a given degree) for each degree may constitute a good set of coarse-grained observables for our problem. Given this knowledge, the coarse timestepper consists of the following fundamental steps (see also figure 1):

(a) Prescribe a coarse-grained initial condition (e.g. values of the average state for each connectivity degree in the network) $u(t_0)$;

(b) Transform the coarse-grained initial conditions through a lifting operator, say $\mu$, to consistent microscopic ensembles, i.e. get $U(t_0) = \mu\, u(t_0)$;

(c) Evolve these microscopic distributions in time with the detailed simulations and report their values after an appropriately chosen short macroscopic time $T$ (time-horizon).

(d) Compute the corresponding coarse-grained variables by restricting the resulting microscopic distributions using a restricting operator, say $M$, i.e. compute $u(T) = M\, U(T)$.

The above procedure can be considered as a "black box" coarse timestepper of the coarse-grained system observables: a map that, using the detailed simulations, and given an initial coarse-grained state $u$ at time $t_0$ will report the result of the evolution after a given time-horizon $T$, i.e.:

$u(t_0 + T) = \Phi_T(u(t_0), p)$, where $p$ denotes the systems parameters.

At this point one can

(e) Utilize computational superstructures like the Newton's or Newton- GMRES method [71, 72] as a shell "wrapped around" the coarse timestepper to compute and trace branches of coarse-grained equilibria or periodic solutions (even unstable ones) [73]; other computational superstructures (other "wrappers") e.g. iterative eigensolvers such as Arnoldi's algorithm [74, 75] can also be used to extract information about the stability of the coarse-grained system dynamics. For example coarse-grained (macroscopic) equilibria can be obtained as fixed points of the mapping $\Phi_T$:

$$u - \Phi_T(u, p) = 0$$

The tracing of solution branches past turning points in one parameter space can be obtained by augmenting the above system with the linearized pseudo arc-length condition:

$$N(u, \mu, \mu) = \alpha \cdot (u - u_1) + \beta(p - p_1) - \Delta s = 0,$$

where

$$\alpha \equiv \frac{(u_1 - u_0)^T}{\Delta s}, \quad \beta \equiv \frac{(\mu_1 - \mu_0)}{\Delta s}$$

and $\Delta s$ is the pseudo arc-length continuation step. $(u_0, p_0)$ and $(u_1, p_1)$, represent two already computed solutions. The pseudo arc-length condition constrains the consequent solution $(u, p)$ to lie on a hyperplane perpendicular to the tangent of the bifurcation diagram at $(u_1, p_1)$, approximated through $(a, \beta)$ and at a distance $\Delta s$ from it. The computation of the fixed points can now be obtained using an iterative procedure like the Newton-Raphson technique. The procedure involves the iterative solution of the following linearized system:

$$\begin{bmatrix} I - \dfrac{\partial \Phi_T}{\partial u} & -\dfrac{\partial \Phi_T}{\partial p} \\ a & \beta \end{bmatrix} \begin{bmatrix} \delta u \\ \delta p \end{bmatrix} = \begin{bmatrix} u - \Phi_T(u, p) \\ N(u, p) \end{bmatrix}$$

Note that for the calculation of both the Jacobian $\dfrac{\partial \Phi_T}{\partial u}$ and the derivative $\dfrac{\partial \Phi_T}{\partial p}$, no explicit macroscopic evolution equations are required. They can be approximated numerically by calling the black-box coarse timestepper at appropriately perturbed values of the corresponding unknowns $(u, p)$. Hence, the above framework enables the microscopic simulator to converge to both coarse-grained stable and unstable solutions and trace their locations, i.e. to fulfill tasks that the simulator was not explicitly designed for.

The main assumption behind the coarse timestepper is that a coarse-grained model for the microscopic-scale dynamics exists and closes in terms of (relatively few) coarse-grained variables which serve as the system observables. These are usually some low-order moments of the microscopically evolving distributions. The assumption about the existence of a coarse-grained model implies that the higher order moments of the microscopic distributions become, relatively quickly over the macroscopic time scales, functionals of (become "slaved to") the few lower ones. What a coarse time-stepper does, in fact, is a model reduction "on demand".

The above methodology essentially sidesteps the derivation of a closed form macroscopic description: relatively short and appropriately initialized short runs of the microscopic simulation provide the necessary closure (refer to [63, 66, 67] for more detailed discussions). The construction algorithms of the lifting and restricting operators for networks that can be described by the degree distribution are introduced in the following section.

### 3.1 Lifting and Restriction operators for a certain class of Social Networks under the Equation-Free Approach

Here we consider the case of undirected graphs for which their connectivity structure is described according to the degree distribution.

Let $U(x)$ denote the distribution function of the microscopic variables over the set of microscopic coordinates $x$. The key idea underlying coarse-graining and model reduction is that certain "fast" system observables (e.g. higher-order moments of the evolving microscopic distributions $U$) become rapidly (say, over a few characteristic interaction times in an individual-based simulation) slaved to the "slow" observables (e.g. the lower order moments) $u$. This functional relation can be stated in a general form as

$$u = \int h\left[U(x)\right] dx.$$

Assuming the existence of a separation of time scales (and an associated slaving) in the overall system dynamics, we can proceed to constructing our restriction and lifting operators. We first start by describing the restriction procedure. For the sake of clarity of the presentation we will illustrate both procedures by considering the average value over the microscopic state.

Let the overall network ensemble assembled by $M$ social networks, $G_j = (V_j, E_j)$, where $1 \leq j \leq M$.

We start by averaging the values of the microscopic variables $U(x)$ over each degree and for all networks in the ensemble. We subsequently sort each individual according to its degree and we assign to it the average value of the microscopic variables. Having all individuals sorted and assigned the "new-averaged" values of the microscopic variable, we proceed to the computation of the coarse-grained variables as follows: let $m$ be the number of coarse-grained variables and $K$ be the total number of agents, over our entire network ensemble simulation; then each coarse-grained variable is defined as the average microscopic state over each $K/m$ fraction of the overall agent population, sorted by the degree.

We define the superset of the vertices of all created networks, $V = \bigcup_{j=1}^{M} V_j$. Given $V$ we proceed to construct the subsets of $V$ containing only vertices of the same degree: $v \in D_i \subseteq V \Leftrightarrow \deg(v) = i, \forall v \in V$; the cardinality of each such set is $d_i = |D_i|$.

Each agent $v$ has a microscopic state $U_v(x)$. For each degree $i$ we calculate the average value of the microscopic state $\bar{U}_i(x)$ as

$$\bar{U}_i(x) = \sum_{v \in D_i} U_v(x) / d_i$$

The total agent population size is $K = MN$. We define a new set containing all the agents sorted by their degree $S \subseteq V$, $S = \{v_1, v_2, ..., v_K\}$, $\deg(v_i) \leq \deg(v_l)$, $i < l$

If $m$ is the number of the coarse-grained variables, each coarse-grained variable represents $K/m$ agents. Thus the restriction of the microscopic distribution(s) to the coarse-grained variables (observables) reads:

$$\forall i \in N, 1 \leq i \leq m: \ obs_i = \sum_{q=(i-1)K/m+1}^{iK/m} \bar{U}_{\deg(v_q)} / (K/m), \ v_q \in S$$

The lifting operation consists of the inverse actions. The inputs to the lifting subroutine are the network parameter value (such as the probability $p$ in the discrete geometric distribution) and the instantaneous values of the $m$ coarse-grained variables. We first create a graph by sampling over the probability distribution. Then we sort its $N$ agents based on their degree. Finally, we assign to each agent in the sorted list the microscopic state given by the corresponding coarse-grained variable value.

Consider the $j$-th lifting attempt, giving rise to the social network $G_j = (V_j, E_j)$ in our ensemble of networks. We then construct the set that contains all the nodes of the set $V_j$ sorted by their degree:

$$S_j \subseteq V_j, \ S_j = \{v_1, v_2, ..., v_N\}, \ \deg(v_i) \leq \deg(v_l), \ i < l$$

Finally we assign to each agent in the $j$-th network a microscopic state according to the following relation:

$$\forall v_i \in S_j: \ U_{v_i}(x) = obs_{floor[(i-1)/(N/m)]+1}$$

where $floor(h/b)$ is the integer part of the division $h/b$.

Alternative ways of constructing the lifting and restricting operators based on the cumulative distribution function of the degree distribution [26] can be also considered.

## 3. The Agent-Based Model

For our illustrations we used a simple dynamic agent-based model inspired by the work presented in [36]; we altered their basic model to include a social network structure - whose construction is presented in the following section- and to implement a kind of information propagation through agent interactions. Our model consists of *N* agents that "live" in our artificial social system caricature and a graph that describes the connections between these individuals.

The agents are characterized by a single real variable which quantifies their emotional state, denoted by $x$, where $x \in [-1,1]$. Agents receive information at discrete time instances (Poisson-distributed in time) from the external environment (public information) as well as from their links in their social environment (private information) changing their state according to the following rules:

(1) In the absence of any public or private information arrival during a time interval $\Delta t$, an agent's state decays exponentially towards zero in time as $x(t) = x(0)\exp[-\gamma \Delta t]$, where $\gamma$ is the decay rate (here constant, a model parameter) and $x(0)$ is the state at the beginning of the time interval. Time intervals in our model are defined in terms of the time unit of our system, *timeUnit*.

(2) *Public* information arrives to each agent as "good" or "bad" news, each type according to a Poisson distribution with arrival rates $v^+$ and $v^-$ respectively. When an agent receives good (bad) external news then its emotional state instantaneously jumps by a finite (typically small and constant, a parameter of the model) positive (negative) amount denoted by $\varepsilon^+$ ($\varepsilon^-$).

(3) *Private* information is defined as information obtained through the agent's social environment. The arrivals of such information also follow a Poisson distribution but with rate equal to $k = a*\deg(i)$, where *α* is a scalar (constant, a parameter of our model) denoting the frequency of "encounters" that an agent initiates with another one of its "environment". When an agent initiates an "encounter" with another one the information received (the change in its emotional state) is considered "good" or "bad" (results in a positive change $e^+$ or a negative change $e^-$ respectively) based on a comparison of its current emotional state with the emotional state of the agent encountered *as recorded at the previous observation time step (see below)*. If the initiating agent is at a worse (algebraically smaller) emotional state than the encountered one, then the incoming information to him/her is classified as "good" news; if he/she has a better (algebraically bigger) emotional state, it is classified as "bad" news.

(4) If during the simulations the emotional state of an agent reaches or exceeds one of the boundary values (-1 and 1) then its state "locks" there for the remaining fragment of the reporting horizon $\delta t$, independently of the arrival of new information. By this we wanted to simulate situations where the emotional state "saturates" – the agent has "made up his/her mind".

## 3.1 The construction of the social network

Consider a graph $G = (V, E)$, where $V$ is the set of vertices - which in our model correspond to the $N$ individuals $\{v_i \mid i = 1, ..., N\}$ - and $E$ is the set of edges, the links between the individuals. An edge $e$ is defined by $\{v_i, v_j\}$ where $v_i, v_j \in V$ are the nodes connected by the edge, and will be referred to as $e_{v_i v_j}$. The graph is undirected so that $e_{v_i v_j} \equiv e_{v_j v_i}$; multiple edges and self-loops are precluded, i.e., $\not\exists e_{v_i v_j} \in E$ where $v_i = v_j$. We considered random connected graphs whose degree sequences are sampled from a slightly altered (in terms of the range of $k$) discrete geometric distribution[†] with parameters $p$ and maximum degree $\max Deg$:

$$y = y(k; p) = p(1-p)^k / \sum_{i=1}^{\max Deg} p(1-p)^i$$

where $y$ is the probability of a vertex with degree $k \geq 1$.

Here $\max Deg$ is set to 140; this choice was based on the so-called *Dunbar's number* (see [77, 49, 78]); $1 \leq k \leq 140$ so that the degree $\deg(v_i)$ of each node $v_i$ lies between $1 \leq \deg(v_i) \leq 140$. At this point we should note that for our illustrations we used a geometric distribution in order to create a right-hand skewed degree of vertices as this characteristic pertains to most real-world networks [26]. In fact we could use any other distribution, skewed or not.

Our main goal is to examine the dynamic behavior of the system described above as the graph topology (in terms of the parameter $p$) varies.

More details on the construction of a connected random graph with a given degree probability distribution are given in the Appendix.

## 4. Simulation Results and Discussion

### 4.1 Model Dynamics

Simulation results were obtained with the following parameter values:
$\delta t = 0.25$, $\gamma = 1$, $v^+ = 20$, $v^- = 20$, $a = 2$, $\varepsilon^+ = 0.075$, $\varepsilon^- = 0.072$, $e^+ = 0.033$,
$e^- = 0.035$, $N = 20000$.
Figure 2 depicts the temporal evolution of the averaged (over $l = 1920$ different network realizations over all nodes with the same degree, and over all degrees) emotional state at each reporting time-instant. Depending on the value of the probability $p$ of the geometric discrete distribution, our network is capable of exhibiting interesting nonlinear behavior.

---

[†] From now on we will refer to it as geometric discrete distribution.

For large values of *p* the average emotional state of the social network converges to a single stationary point, independently of the initial conditions of the emotional state of our agents (see figures 2a-b). As the value of *p* decreases (making our graph denser, as far as the number of edges is concerned) - and now depending on the initial conditions - a second stable stationary point is revealed (figure 2c). Further decrement of the value of *p* - depending again on the initial conditions - brings up a new stationary point (figure 2d) making the number of stable stationary average states of the network equal to three. Eventually, when *p* decreases beyond a critical value, the middle stable stationary point disappears (figure 2e-f). The above results are indicative of the existence of several critical parameter values (values of the discrete geometric distribution parameter *p*) marking qualitative transitions in the dynamic behavior of our system - and as the bifurcation theory dictates - the onset of unstable coarse-grained states. Based on the results of our temporal simulations, these three critical values lie in the ranges:

$$0.084 < p^1_{critical} < 0.089,\ 0.079 < p^2_{critical} < 0.084,\ 0.03 < p^3_{critical} < 0.04$$

The temporal simulations shown at figure 2 indicate that networks with the same degree distribution (and even further, networks with degrees sampled from the discrete geometric distribution with the same probability *p*) can have one, two or even three stable stationary states and unstable ones (which are unreachable through long-temporal simulations). To systematically study the nonlinear behavior of the system we employed the "Equation-Free" approach to construct the coarse-grained bifurcation diagram.

As we noted in the previous sections the key idea underlying the Equation-Free approach is that the higher-order moments of the evolving distribution evolve very fast, compared to the macroscopic time-horizon *T*, becoming quickly functionals of a few lower-order moments. In other words, the system evolves quickly to a low-dimensional, attracting slow manifold. In order to provide computational support for this slaving we performed various temporal runs near the stationary solutions of all three apparently stable solution branches involving the first moments of the underlying microscopic distribution, namely the overall network average emotional state $\bar{x}$ and the norm $\|\bar{s}\|$ of the vector of variances of the average emotional state for each degree.

These variances are computed as follows: assume that we have created *l* consistent networks: $G_j = (V_j, E_j)$, where $1 \leq j \leq l$ with *N* agents per network and a maximum degree equal to $\max Deg$. Based on the sets $V_j$ for each social network *j* we create the following new sets:

$D_{ij} \subseteq V_j$ where $\forall v \in V_j$ if $\deg(v) = i \iff v \in D_{ij}$

Each agent has an emotional state is denoted by $x(v)$, where *v* denotes the individual. For each network *j* and each degree *i* the average emotional state and the corresponding variance are $\bar{x}_{ij}$ and $s_{ij}$ respectively. These are computed by the following relations:

$$\bar{x}_{ij} = \sum_{v \in D_{ij}} x(v) / d_{ij},\text{ where } d_{ij} = |D_{ij}| \text{ and}$$

$$s_{ij} = (\sum_{v \in D_{ij}} (x(v) - \bar{x}_{ij})^2) / d_{ij}.$$

For each $j$ we compute the network average emotional state $\bar{x}_j$, reading

$$\bar{x}_j = \sum_{v \in V_j} x(v) / |V_j|.$$

Over the set of networks the resulting overall network average emotional state is

$$\bar{x} = (\sum_{j=1}^{l} \bar{x}_j) / l.$$

We construct a vector $\bar{s} = \{\bar{s}_1, \bar{s}_2, ..., \bar{s}_{\max Deg}\}$ where

$$\bar{s}_i = (\sum_{j=1}^{l} s_{ij}) / l, \text{where } 1 \leq i \leq \max Deg$$

which is the average variance of the emotional state for each degree over the entire set of social networks that we have created and simulated.

All the phase portraits (see figure 3 a-f) depict four transients, each initialized with all individuals with the same initial emotional state (thus having a zero variance for each degree). A fifth transient simulation, initialized based on the statistics of a stable stationary point of a different set of networks - sampled with different $p$ from the discrete geometric distribution - is also depicted. Initialization of this transient involves application of the restriction and then of the lifting operations. Clearly, as is shown in all phase portraits, after a relatively short transient, the norm $\|\bar{s}\|$ of the vector of variances becomes a function of $\bar{x}$: it evolves quickly towards, and subsequently on a *slow manifold* parameterized by the first-order moment $\bar{x}$ of the underlying distribution. The alternatively initialized fifth transient is the middle one in each panel; it evolves (even faster) to the same slow manifold.

**4.2 The coarse-grained bifurcation diagram and the associated stability analysis**
Having described the choice of coarse-grained variables as well as our restriction and lifting operations, we are ready to proceed to the construction of the coarse-grained bifurcation diagram for our system. The stationary states on the coarse-grained bifurcation diagram have been obtained as fixed points of the agent-based timestepper averaged over 1920 realizations (this particular number, 6 x 320, was dictated by our six-node cluster), by wrapping Newton-GMRES [71], as a shell, around the coarse timestepper. Continuation around the coarse-grained turning points is accomplished by coupling the fixed point algorithm with pseudo-arc-length continuation. The critical eigenvalues are computed by implementing an Arnoldi iterative eigensolver which was again wrapped around our coarse timestepper. In our computations we used eighty coarse-grained observables ($m = 80$ in restriction and lifting operators) and the time horizon $T$ was set to seven $\delta ts$. The coarse-grained bifurcation diagram constructed through this approach is shown in figure 4. Two sets of branches (which we will refer to as the "upper" and the "lower" ones) are found. The five insets show the average emotional state for each degree at the five stationary points (marked a-e) coexisting at approximately same $p$ value. The insets above the bifurcation diagram depict representative connectivity degree probability distributions at four different values of $p$. As it is shown, steeper distribution curves (indicating that the majority of the nodes in the network have a rather small number of links) drive the system to unique stable stationary states. On the other hand, looser distribution curves (indicating that a large percentage of nodes have a large number of links) result to the emergence of multiple stationary states; depending on the initial conditions the system reaches one of them.

To perform linearized stability analysis of the obtained coarse-grained solutions we numerically approximated the dominant/critical eigenvalues of the macroscopic dynamics using the Arnoldi iterative eigensolver. Figure 5 shows the variation of the largest eigenvalue as the bifurcation parameter $p$ varies along the upper solution branches in the coarse-grained bifurcation diagram. Figure 6 shows the five largest coarse eigenvalues at two points, just before (figure 6a) and just after (figure 6b) the "1" saddle-node bifurcation, in the complex plane. Figure 7 shows the largest eigenvalue of the coarse-grained linearization as the bifurcation parameter $p$ varies along the lower part of branches of the coarse-grained bifurcation diagram. Figure 8 shows the five largest eigenvalues at four points, just before (figure 8a,c) and just after (figure 8b,d) the two saddle-node bifurcations marked as "2" and "3" respectively.

The obtained state distributions and coarse-grained stability results are consistent with temporal agent-based simulations.

## 5. Discussion and Conclusions

We showed how the so called "Equation-Free" approach for complex/multiscale computations can be exploited to systematically analyze the coarse-grained dynamics of a certain class of social networks. This was accomplished through the design of computational experiments using the detailed, agent-based simulation as a black-box coarse timestepper. For our illustrations we used a computational caricature of a mimesis-like propagation process in a simple social network of individuals. The agent-based simulations show some interesting non-linear behavior including apparent phase transitions and multiplicity of stationary states. We constructed coarse-grained bifurcation diagrams with respect to a parameter dictating the degree distribution of the networks and analyzed the (coarse) stability of the stationary solutions. At this point we should note that the identification of multiple stable equilibria is in agreement with empirical evidence and studies of behavioral and financial processes as well as other instances of complex systems (such as information propagation in networks and neuroscience) [79, 23, 80].

Further research could proceed along several directions including the study of "rare event" transitions at the neighborhood of coarse-grained bifurcation points, and the use of Diffusion Map techniques [70, 81] that can be used to both extract the right variables and to efficiently extract information from more complicated and thus more realistic agent-based social network simulations.

One can also envision using such coarse-grained techniques to enhance the study the dynamics of complex network/population models which incorporate decision making based on agent-based sociological models (econophysics [82, 83]), to explore strategy formation, formulate optimal control policies, and more generally, in order to prescribe/control the system (network) dynamics. Traditional numerical tools for performing such tasks in continuum models are, in principle, not applicable when the model is a detailed agent-based one. "Equation-Free" techniques hold the promise of linking agent-based codes with traditional continuum numerical analysis; the compactness of the coarse descriptions and the fast and quantitative extraction of macroscopic information from realistic fine scale codes may well enhance the use of agent-based models for understanding, analyzing and even controlling certain social phenomena.

**Acknowledgements:** This work was partially supported by the State Scholarships Foundation of Greece (IKY) (ACT, GVB) and the National Technical University of Athens through the Basic Research Program ''Constantin Caratheodory'' (ACT, CIS). IGK acknowledges partial support by the US NSF and DTRA.

REFERENCES

[1] D.L. DeAngelis, K.A. Rose, and M.A. Huston, *Frontiers in Mathematical Biology.* (Springer, Berlin, 1994), p. 290.
[2] V. Grimm, Ecol. Model. **115**, 129 (1999).
[3] E. Bonabeau, M. Dorigo, and G. Theraulaz, Nature **406**, 39 (2000).
[4] I. M. Longini, P. E. Fine, and S. B. Thacker, Am. J Epidemiol. **123**, 383 (1986).
[5] S. H. Eubank, V. S. A. Guclu, M. Kumar, M. V. Marathe, A. Srinivasan, Z. Toroczkai, and N. Wang, Nature **429**, 180 (2004).
[6] N. M. Ferguson, D. A. T. Cummings, S. Cauchemez, C. Fraser, S. Riley, A. Meeyai, S. Iamsirithaworn, and D. S. Burke, Nature **437**, 209 (2005).
[7] M. Raberto, S. Cincotti, S.M. Focardi, and M. Marchesi, Physica A **219**, 319 (2001).
[8] G. Iori, Journal of Economic Behavior & Organization **49**, 269 (2002).
[9] S. Wang, and C. Zhang, Physica A **354**, 496 (2005).
[10] E. Samanidou, E. Zschischang, D. Stauffer, and T. Lux, Rep. Prog. Phys. **70**, 409 (2007).
[11] X. Liu, S. Shirley Gregor, and J. Yang, Physica A **387**, 2535 (2008).
[12] J. Dijkstra, A.J. Jessurun, and H.J.P. Timmermans, *Pedestrian and Evacuation Dynamics* (Springer-Verlag, Berlin, 2001) p. 173.
[13] D. Helbing, Rev. Modern. Physics **73**, 1067 (2001).
[14] N. Julka, R. Srinivasan, and I. Karimi, Computers and Chemical Engineering **26**, 1755 (2002).
[15] B. Raney, N. Cetin, A. Völlmy, M. Vrtic, K. Axhausen, and K. Nagel, Networks and Spatial Economics **3**, 23 (2003).
[16] M. Balmer, K. Nagel, and B. Raney, Journal of Intelligent Transportation Systems **8**, 205 (2004).
[17] H. Levine, and W. J. Rappel, I. Cohen, Phys. Rev. E **63**, 017101 (2000).
[18] Y. Liu, and K. M. Passino, IEEE Trans. Autom. Contr. **49**, 30 (2004).
[19] E. Bonabeau, PNAS **99**, 7280 (2002).
[20] A. M. Khulood, and A. Z. Raed, Information and Software Technology **49**, 695 (2007).
[21] B.Bollobàs, *Graph theory* (Springer, New York, 1979).

[22] S. H. Strogatz, Nature **410**, 268 (2001).
[23] D. S. Callaway, J. E. Hopcroft, J. M. Kleinberg, M. E. J. Newman and S. H. Strogatz, Physical Review E **64**, 041902 (2001).
[24] M. E. J. Newman, D. J. Watts and S. H. Strogatz, PNAS **99**, 2566 (2002).
[25] R. Albert and A. L. Barabasi, Reviews of Modern Physics **74**, 47 (2002).
[26] M. E. J. Newmann, SIAM review **45** (2), 167 (2003).
[27] G. Szabó, and G. Fàth, Physics Reports **446**, 97 (2007).
[28] M. Boots, and A. Sasaki, Proc. R. Soc. B **266**, 1933 (1999).
[29] M. Bogunà and R. Pastor-Satorras, Phys. Rev. E **66,** 047104 (2002),
[30] M. J. Kelling, and K. T. D. Eames, J. R. Soc. Interface **2**, 295 (2005).
[31] M. J. Keeling, Theor. Popul. Biol. **67**, 1 (2005).
[32] K. Daniel, D. Hirshleifer, and S. H. Teoh, Journal of Monetary Economics **49**, 139 (2002).
[33] N. Barberis, *Elsevier Handbook of the Economics of Finance* (Elsevier North-Holland, 2003) Vol. 1B, Chapter 18, p. 1052.
[34] J. R. Ritter, Pacific-Basin Finance Journal **11**, 429 (2003).
[35] L. Palatella, J. Perellò, M. Montero, and J. Masoliver, Physica A **355**, 131 (2005).
[36] A. Omurtag, and L. Sirovich, Journal of Economic Behavior & Organization **61**, 562 (2006).
[37] D. Hirshleifer, A. Subrahmanyam, and S. Titman, Journal of Financial Economics **81**, 311 (2006).
[38] K. Kim, S-Y. Kim, and D-H. Ha, Computer Physics Communications **177**, 184 (2007).
[39] V. Boginski, S. Butenko, and P. M. Pardalos, Computers & Operations Research **33**, 3171 (2006).
[40] J. Reichardt and D.R. White, Eur. Phys. J. B **60**, 217 (2007).
[41] H. Oliver, S. Michael, and S. Markus, Journal of Economic Interaction and Coordination **1**(3), 59 (2008).
[42] Clifford, P., and A. Sudbury, Biometrika **60**(3), 581 (1973).
[43] R. Holley, and T. Liggett, Ann. Probab. **3**(4), 643 (1975).
[44] L. Steels, and J.-C. Baillie, Trends Cogn. Sci. **7**(7), 308 (2000).
[45] C. J. Tessone, and R. Toral, Physica A **351,** 106 (2005).
[46] Claudio Castellano, Santo Fortunato, and Vittorio Loreto, *Statistical physics of social dynamics,* Eprint arXiv: 0710.3256
[47] R. Lambiotte, M. Ausloos and J. A. Holyst*,* Phys. Rev. E **75,** 030101(R) (2007).
[48] R. Lambiotte, S. Thurner and R. Hanel*,* Phys. Rev. E **76,** 046101 (2007).
[49] R.I.M. Dunbar, Behavioral and Brain Sciences **16**, 681 (1993).
[50] R. Axelrod, *The Complexity of Cooperation* (Princeton University Press, Princeton, 1997).
[51] Claudio Castellano, Matteo Marsili, and Alessandro Vespignani, Physical Review Letters **85**, 3536 (2000).
[52] S. N. Dorogovtsev and J. F. F. Mendes, Proc. Roy. Soc. London Ser. B **268**, 2603 (2001).
[53] Vittorio Loreto and Luc Steels, Nature Physics **3**, 758 (2007).
[54] D. Helbing and P. Molnár, Physical Review E **51**, 4282 (1995).

[55] R. L. Hughes, Math. Comp. Simul. **53**, 367 (2000).
[56] D. Helbing, I. Farkas, and T. Vicsek, Nature **407**, 487 (2000).
[57] D.R. Parisi and C.O. Dorso, Physica A **354**, 606 (2005).
[58] X. Guardiola, A. Diaz-Guilera, C. J. P´erez, A. Arenas, and M. Llas, Phys. Rev. E **66**, 026121 (2002).
[59] M. Llas, P. M. Gleiser, J. M. L´opez, and A. Diaz-Guilera, Phys. Rev. E **68**, 066101 (2003).
[60] F. Wu, and B. A. Huberman, Proc. Natl. Acad. Sci. USA **104**(45), 17599 (2007).
[61] Weidlich, W., *Sociodynamics: A Systematic Approach to Mathematical Modelling in Social Sciences* (Taylor and Francis, London, UK, 2002)
[62] C. Theodoropoulos, Y. H. Qian, and I. G. Kevrekidis, PNAS **97**, 9840 (2000).
[63] C. W. Gear, I. G. Kevrekidis and C. Theodoropoulos, Computers and Chemical Engineering **26**, 941 (2002).
[64] A. Makeev, D. Maroudas, and I. G. Kevrekidis, Journal of Chemical Physics **116**, 10083 (2002).
[65] O. Runborg, C. Theodoropoulos and I. G. Kevrekidis, Nonlinearity **15**, 491 (2002).
[66] C. I. Siettos, M. Graham and I. G. Kevrekidis, Journal of Chemical Physics **118**, 10149 (2003).
[67] I. G. Kevrekidis, C. W. Gear, J. M. Hyman, P. G. Kevrekidis, O. Runborg, and C. Theodoropoulos, Comm. Math. Sciences **1**(4), 715 (2003).
[68] I. G. Kevrekidis, C. W. Gear and G. Hummer, AI.Ch.E Journal **50**, 1346 (2004).
[69] W.W. Cooley and P.R. Lohnes, *Multivariate Data Analysis* (John Wiley & Sons, Inc., New York, 1971).
[70] R.R. Coifman, S. Lafon, A.B. Lee, M. Maggioni, B. Nadler, F. Warner and S.W. Zucker, PNAS **102**, 7426 (2005).
[71] C.T. Kelley, *Iterative methods for linear and nonlinear equations* (SIAM, Philadelphia, 1995).
[72] C.T. Kelley, I.G. Kevrekidis and L. Qiao, SIAM Journal of Scientific Computing **19**, 1188 (2004).
[73] M. Kavousanakis, L. Russo, C.I. Siettos, A. G. Boudouvis and G. C. Georgiou, Journal of Non-Newtonian Fluid Mechanics **151**, 59 (2008).
[74] Y. Saad, *Numerical methods for large eigenvalue problems* (Manchester University Press: Oxford-Manchester, 1992).
[75] K.N. Christodoulou and L.E. Scriven, SIAM Journal of Scientific Computing **3**, 355 (1998).
[76] R. J. Shiller, *Elsevier Handbook of Macroeconomics* (Elsevier 1999), Volume 1, chapter 20, p. 1305.
[77] J. J. Edney, J. of Comm. Psych. **9** (1981a).
[78] M. Gladwell, The Tipping point - how little things make a big difference (Little, Brown and Company, 2000).
[79] R. Topol, Economic Journal **101**, 786 (1991).
[80] C. A. Jones, Connecting blazons and neurons, Science **319**, 35 (2008)
[81] Kolpas, A., J. Moehlis, and I. G. Kevrekidis, *Proc. Nat. Acad. Sci. USA* **104**, 5931 (2007)

[82] Rosario N. Mantegna, H. Eugene Stanley, An Introduction to Econophysics: Correlations and Complexity in Finance, Cambridge University Press (Cambridge, 1999).
[83] B. K. Chakrabarti, A. Chakraborti, A. Chatterjee, *Econophysics and Sociophysics : Trends and Perspectives* (Wiley-VCH, Berlin 2006).
[84] M. Molloy and B. Reed, Random Structures & Algorithms **6**, 161 (1995).

**Appendix: The construction of a connected random graph with a given degree probability distribution without self-loops and multiple edges**

We constructed the graph using a modified version of the algorithm described in [84] in order to create a network which precludes the presence of self-loops, multiple edges and ensures that there is at least one path connecting any two nodes of the network. We should note that the algorithm as described in [84] does not guarantee a connected graph and the occurrence of self-loops and multiple edges is not ruled out. The modified algorithm reads as follows:

The degree $\deg(v_i)$ of a vertex $v_i$ of a simple graph $G(V,E)$ is defined as

$$\deg(v_i) = \left| e_{v_i v_j} \right| \quad \forall v_i \in V, \text{ where } e_{v_i v_j} \in E \text{ and } v_j = v_i \text{ or } v_l = v_i.$$

a) For each node choose randomly a degree according to the given discrete geometric distribution. In an undirected graph the sum of the node degrees should be even, since the number of edges is an integer equal to $(\sum_{i=1}^{N} \deg(v_i))/2$. To satisfy this constraint we may need to repeatedly randomly select the degree of the last node so that the sum becomes even. An undirected connected graph has a minimum of $N-1$ edges, where $N$ is the number of vertices. If by sampling there are less than the minimum needed edges we resample the probability distribution.

b) We then randomly connect the vertices of our graph in such a way that there are no two vertices connected by more than one edge and no self - loops.

c) After that, as we want to construct a one component graph, we identify the components of our graph, count them, we identify the edges that participate in circuits[‡] of the components, and we finally link these components. The first two steps are achieved using a modified breadth-first search algorithm with "Last In First Out" buckets, for efficiency, in order to create the spanning trees, thus identifying the components and registering those edges that are not part of the spanning trees (these edges belong to circuits). We list the graph components in descending order of the number of registered edges of each component. Knowing the edges that are part of circuits, we can use them in order to progressively link components without causing any sub-component to become disconnected (the structure of the spanning tree created from each component that has at least a circuit remains unaltered). The steps of the linking algorithm are then:

   a. Register all components in descending order of the number of circuits in a components vector, *component*[], set *componentCounter:=1* and set *numOfComponents* equal to the number of the components of the graph;
   b. Insert all edges that create a circuit (the edges not used in the spanning tree) of the *component*[*componentCounter*] into the *circuit_vector*;
   c. Select one of the edges randomly and remove it from the *circuit_vector;*

---

[‡] A circuit is a walk starting from a node *x* and ending at the same node without having visited any other node more than once or used any edge more than once.

d. *If the number of the circuits of the component[componentCounter+1] are greater than zero then*
   i. select one of the edges that create a circuit and insert all the other edges to the *circuit_vector*;
   ii. Delete the two selected edges and using the four vertices involved create (randomly among the two options available) two new edges, different than the deleted ones;
   iii. Insert one of the new edges (chosen randomly) to the *circuit_vector*;
   *else*
   i. select one of the edges of the component randomly;
   ii. Delete the two selected edges, randomly create two other between the four vertices (different than the deleted ones)
   *endif*;
e. Set *componentCounter:= componentCounter +1*;
f. *If componentCounter < numOfComponents then*
       goto c,
   *endif*;
g. Terminate.

The nature of the problem makes the linking-part of the algorithm simple, in the sense that linking components does not violate - by construction - certain constraints (no self-loops, no multiple edges between nodes) of our graphs. The two largest components of the graph have been constructed randomly and thus a direct connection of them does not insert any bias into the algorithm. Moreover all rewiring parts of the algorithm maintain the degree chosen for each vertex at the first step of the algorithm.

**Figures**

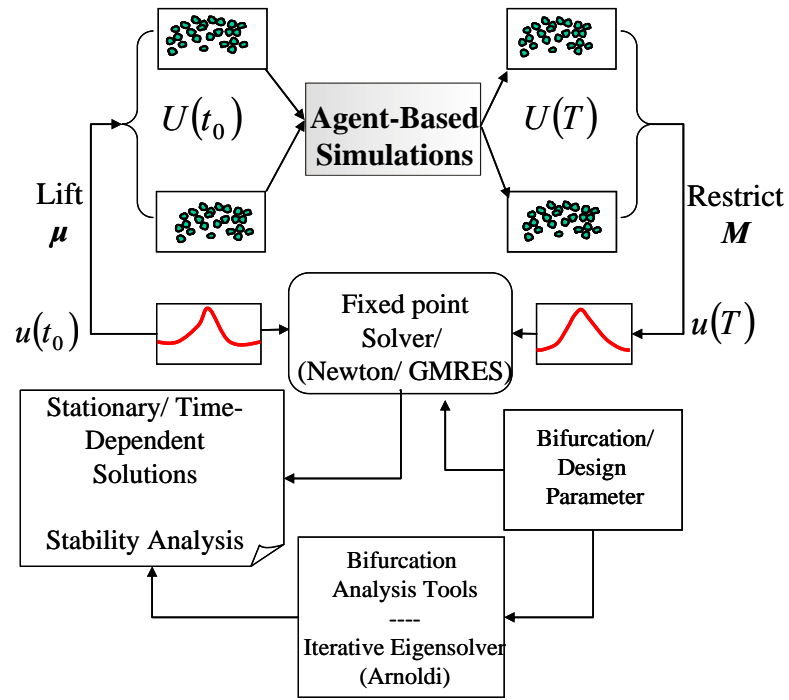

**Figure 1**

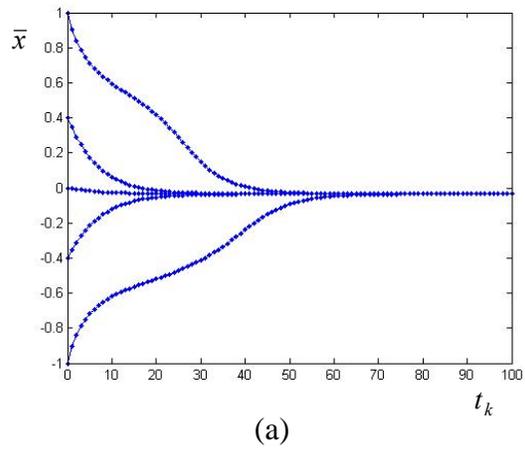
(a)
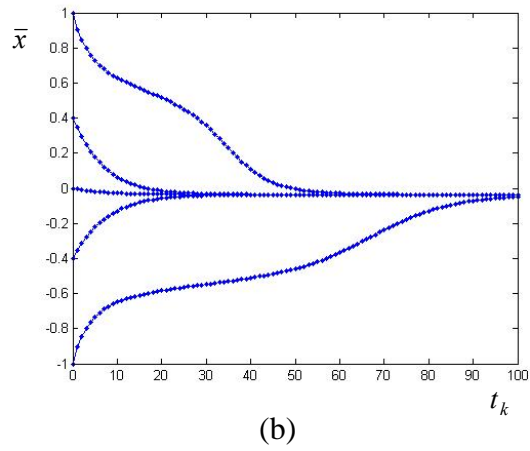
(b)
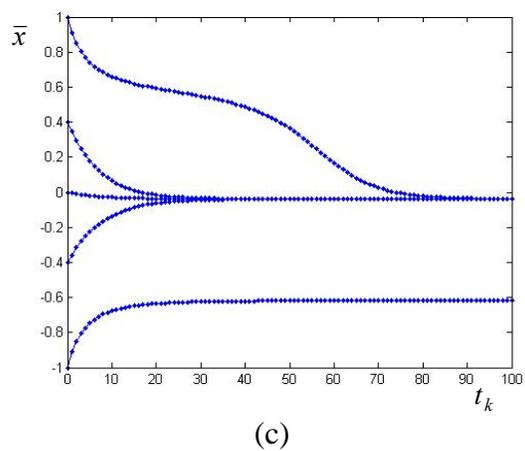
(c)
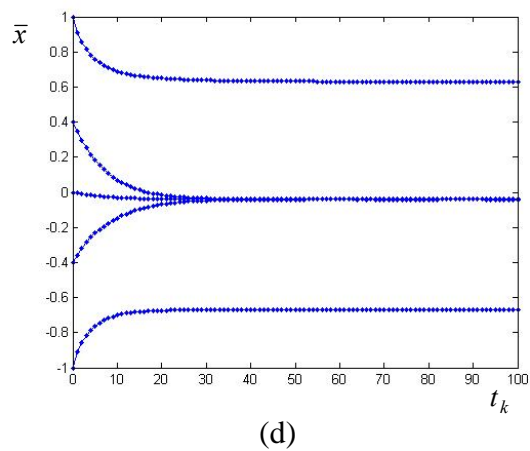
(d)
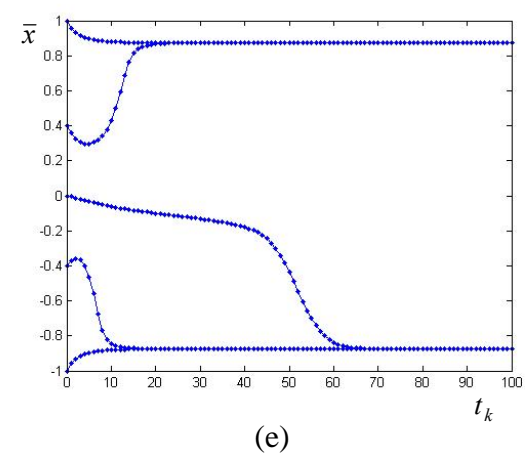
(e)
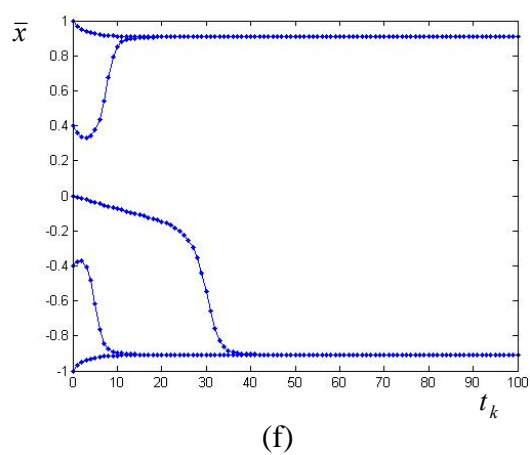
(f)

**Figure 2**

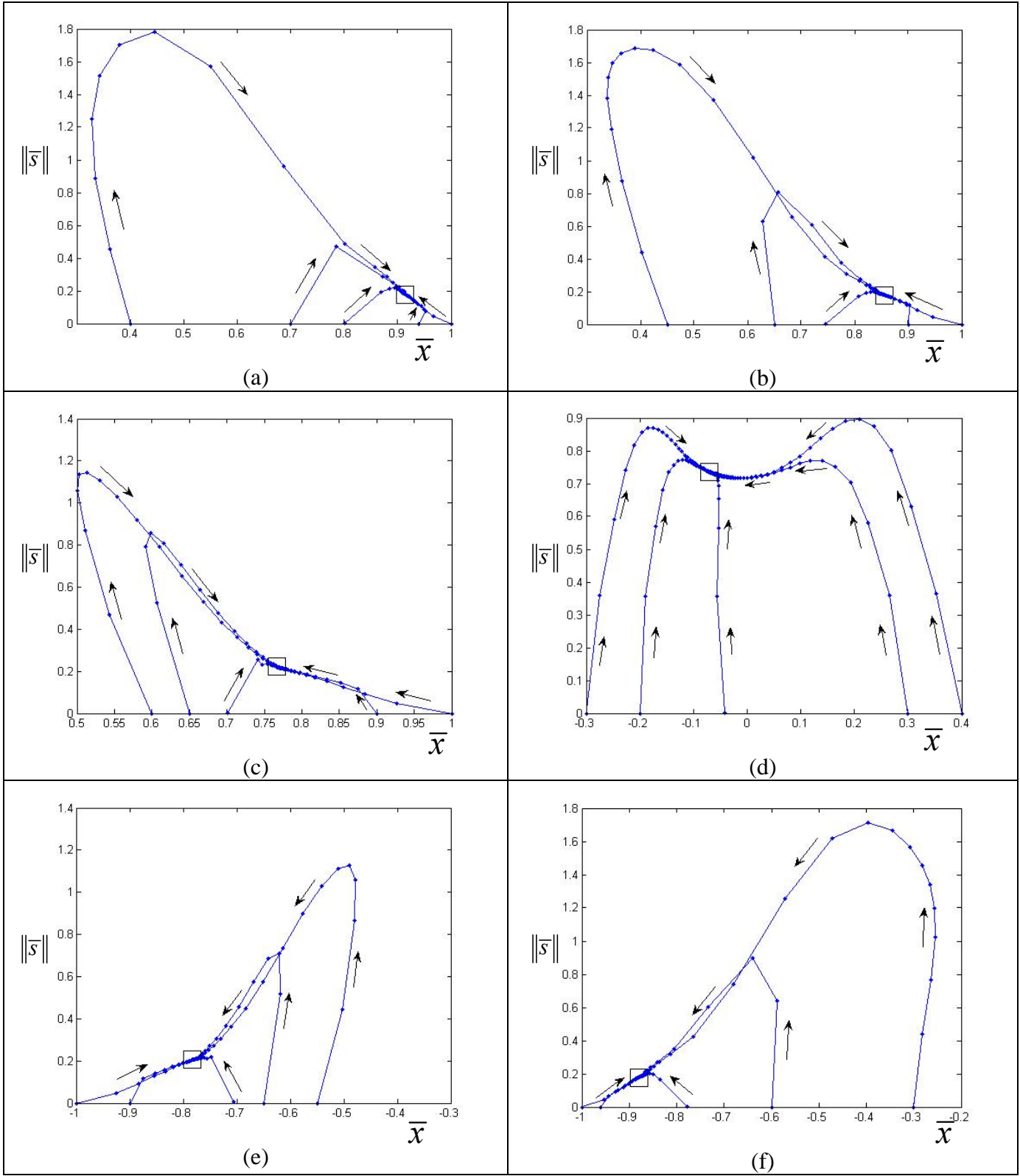

**Figure 3**

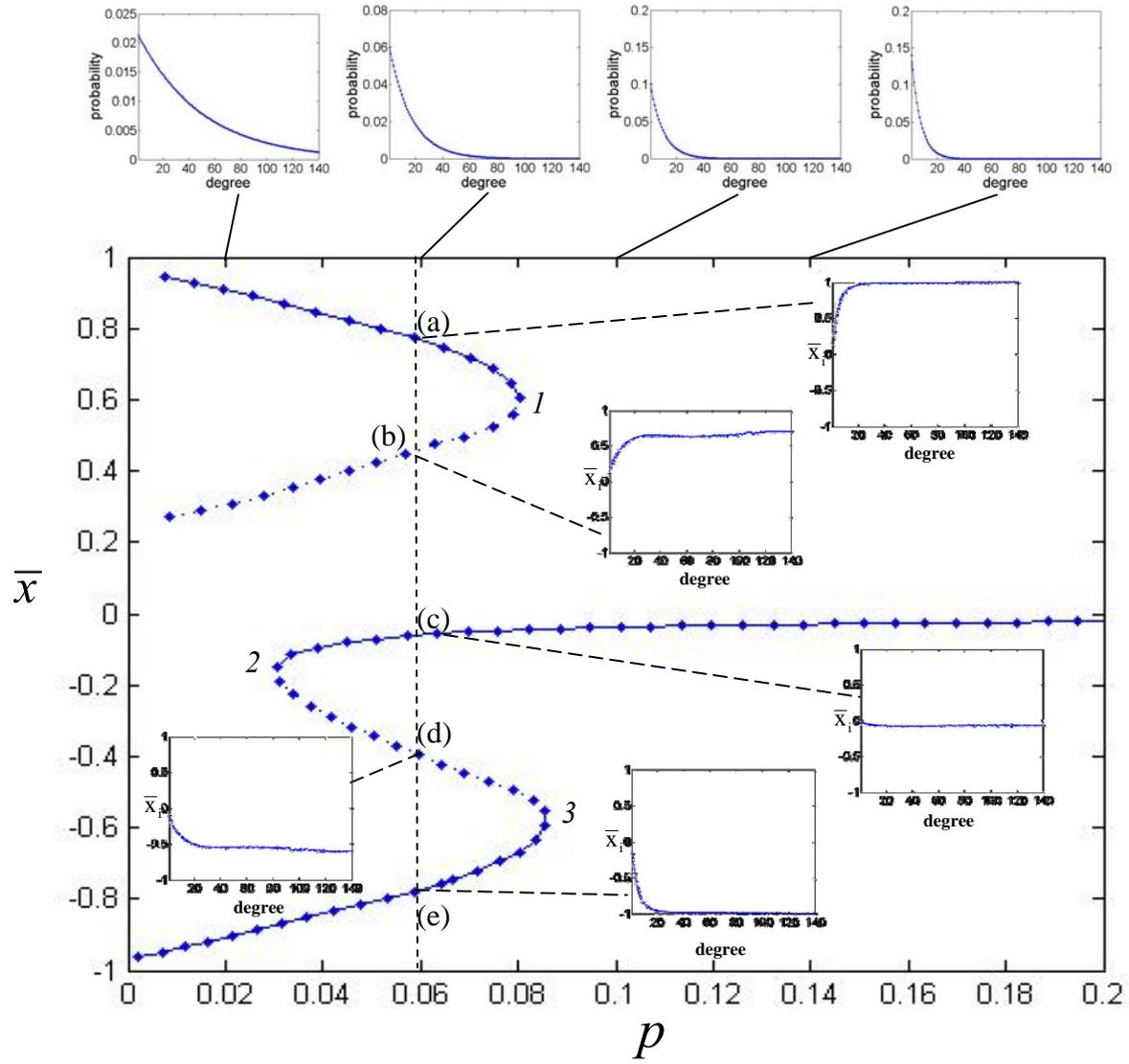

**Figure 4**

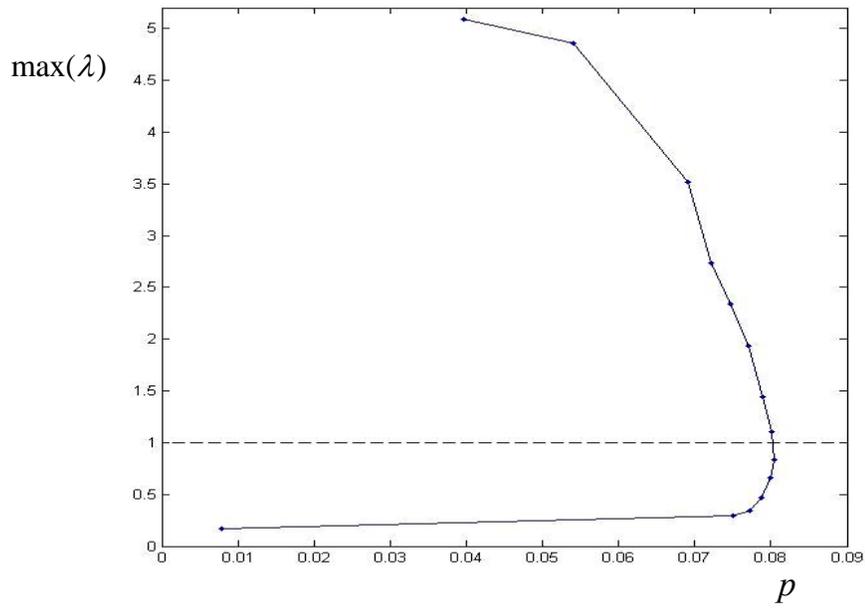

**Figure 5**

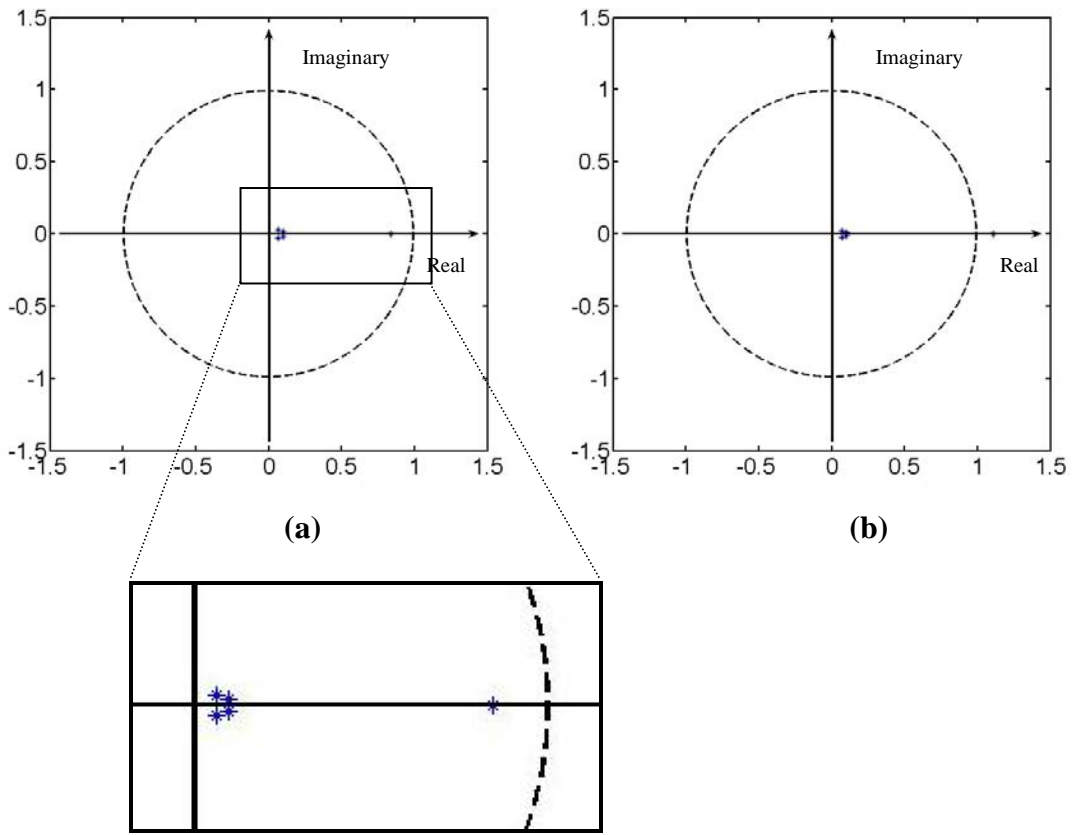

**Figure 6**

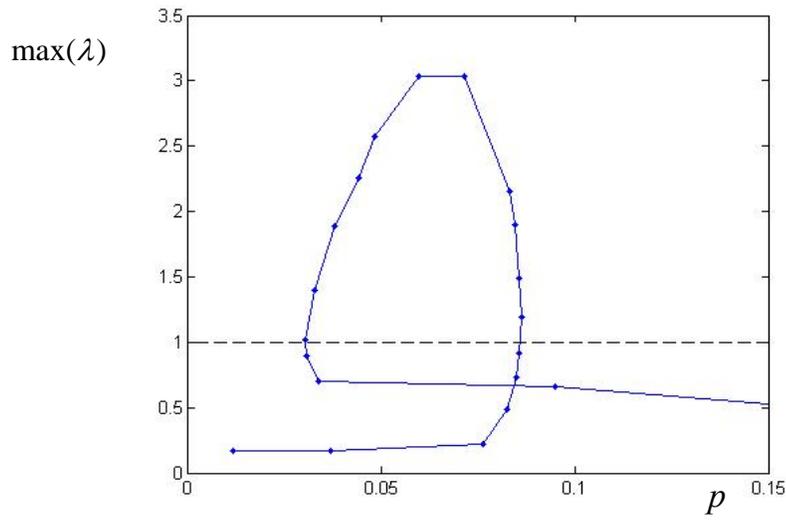

**Figure 7**

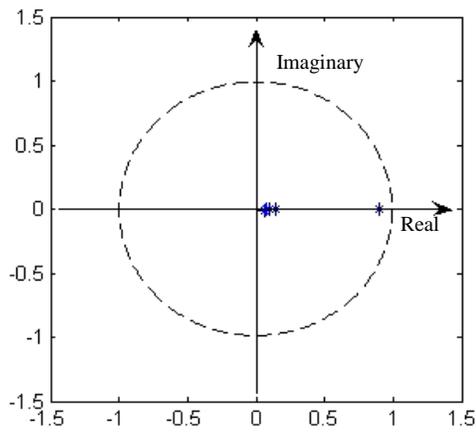
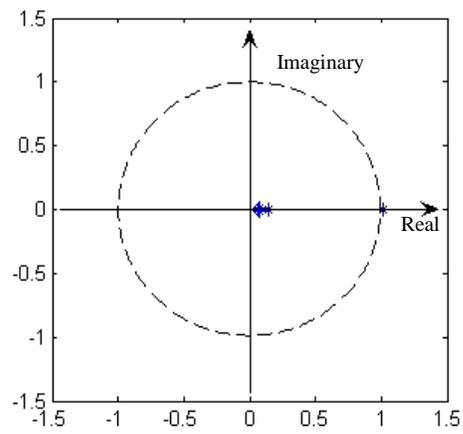

(a)          (b)

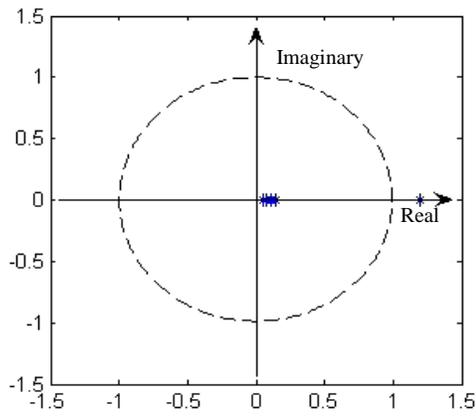
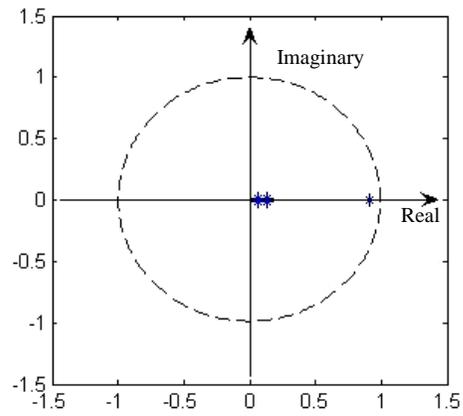

(c)          (d)

**Figure 8**

# Captions of Figures

**Figure 1.** A schematic of the Equation-Free approach.

**Figure 2.** Temporal simulations depicting the evolution of the network average (over 1920 independent realizations) emotional state starting from different initial conditions as the value of the parameter $p$ decreases: (a) $p = 0.094$, (b) $p = 0.089$, (c) $p = 0.084$, (d) $p = 0.079$, (e) $p = 0.03$, (f) $p = 0.02$. A maximum of three stable stationary solutions is possible.

**Figure 3.** Phase portraits in terms of $\bar{x}$ (the overall average network emotional state, the first moment) and the norm $\|\bar{s}\|$ of the vector of variances of the average emotional state for each degree (a higher-order feature) for (a) $p = 0.0197$, (b) $p = 0.0384$, (c) $p = 0.0589$, (d) $p = 0.0572$, (e) $p = 0.0585$, (f) $p = 0.0316$. Notice the fast evolution of all plotted transients to a one-dimensional slow manifold, apparently well-parameterized by $\bar{x}$.

**Figure 4.** Coarse-grained bifurcation diagram, plotted in terms of the overall network average emotional state $\bar{x}$ vs. $p$. It has been obtained by wrapping a Newton-GMRES iterative solver around the coarse timestepper; five coexisting stationary states, for approximately same $p$ value, are marked (a)-(e). The insets show the average emotional state ($\bar{x}_i$, ordinate) for each degree (abscissa) at each of these stationary points, whether stable or unstable. Representative degree probability distributions at four distinct values of p are included above the bifurcation diagram.

**Figure 5.** The leading eigenvalue of the linearized coarse-grained dynamics vs. the bifurcation parameter $p$; it was computed by implementing Arnoldi's algorithm around the coarse timestepper along the "upper" solution branches of figure 3.

**Figure 6.** Five largest eigenvalues of the linearized coarse-grained dynamics along the upper solution branches of the coarse-grained bifurcation diagram (figure 3). (a) just before the "1" saddle-node bifurcation, on the stable branch of coarse-grained equilibria for $p= 0.0804$ and (b) just after the "1" saddle-node bifurcation on the lower-lying, unstable branch for $p= 0.0801$. The blowup in (a) is included to clearly show the four eigenvalues forming an inner "cluster".

**Figure 7.** The leading eigenvalue of the linearized coarse-grained dynamics vs. the bifurcation parameter along the lower solution branches of the coarse-grained bifurcation diagram (figure 3).

**Figure 8.** Five largest eigenvalues of the linearized coarse-grained dynamics along the lower solution branches of the coarse-grained bifurcation diagram (figure 3): (a) just before the "2" saddle-node bifurcation on the branch of coarse-grained stable equilibria for $p= 0.0305$, (b) just after the "2" saddle-node bifurcation on the branch of coarse-grained unstable equilibria for $p= 0.0303$, (c) just before the "3" saddle-node bifurcation on the branch of coarse-grained unstable equilibria for $p= 0.0861$, (d) just after the "3" saddle-node bifurcation on the branch of coarse-grained stable equilibria for $p= 0.0857$.